\documentclass[10pt,aps,prd,superscriptaddress,nofootinbib,nobibnotes,longbibliography,floatfix,twocolumn]{revtex4-2}

\usepackage{bm}
\usepackage{mathtools,
amsmath,
amssymb,
amsfonts,
mathrsfs,
chngcntr,
multirow}

\let\cc\corresponds
\let\corresponds\relax
\usepackage{mathabx}
\let\corresponds\cc

\usepackage[utf8]{inputenc}
\usepackage[T1]{fontenc}
\usepackage[dvipsnames]{xcolor}
\usepackage[unicode]{hyperref}
\hypersetup{colorlinks=true, citecolor=MidnightBlue,
            linkcolor=MidnightBlue, urlcolor=MidnightBlue, linktocpage=true}
\usepackage[normalem]{ulem}
\usepackage{orcidlink}
\usepackage[capitalize]{cleveref}
\usepackage{float}
\usepackage{comment}

\usepackage{orcidlink}

\newcommand{\orcid}[1]{\href{https://orcid.org/#1}{\includegraphics[width=10pt]{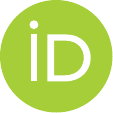}}}

\newcommand{\qm}[1]{``#1''}
\newcommand{\dd}{{\rm d}}

\begin{document}

\title{Parametrized beyond-Teukolsky framework in the time domain}

\author{Ciro De Simone \orcid{0009-0004-0610-1686}}
\email{ciro.desimone@unina.it}
\affiliation{Dipartimento di Fisica \qm{E. Pancini}, Universit\'a di Napoli \qm{Federico II}, Complesso Universitario di Monte S. Angelo, Via Cinthia Edificio 6, I-80126 Napoli, Italy}
\affiliation{Istituto Nazionale di Fisica Nucleare, Sezione di Napoli, Complesso Universitario di Monte S. Angelo, Via Cinthia Edificio 6, 80126 Napoli, Italy}

\author{Sebastian H.\,V\"olkel \orcid{0000-0002-9432-7690}}
\email{sebastian.voelkel@uni-tuebingen.de}
\affiliation{Theoretical Astrophysics, IAAT, University of T\"ubingen, D-72076 T\"ubingen, Germany}

\author{Kostas D.\,Kokkotas \orcid{0000-0001-6048-2919}}
\email{kostas.kokkotas@uni-tuebingen.de}
\affiliation{Theoretical Astrophysics, IAAT, University of T\"ubingen, D-72076 T\"ubingen, Germany}

\author{Salvatore Capozziello \orcid{0000-0003-4886-2024}}
\email{capozziello@na.infn.it}
\affiliation{Dipartimento di Fisica \qm{E. Pancini}, Universit\'a di Napoli \qm{Federico II}, Complesso Universitario di Monte S. Angelo, Via Cinthia Edificio 6, I-80126 Napoli, Italy}
\affiliation{Istituto Nazionale di Fisica Nucleare, Sezione di Napoli, Complesso Universitario di Monte S. Angelo, Via Cinthia Edificio 6, 80126 Napoli, Italy}
\affiliation{Scuola Superiore Meridionale,  Via Mezzocannone 4, I-80134 Napoli, Italy}

\begin{abstract}
Modifications to General Relativity can significantly alter the perturbative response of black holes, leaving imprints on quasinormal-mode spectra, waveform amplitudes and phases, and late-time tails. The parametrized beyond-Teukolsky framework was introduced to capture possible deviations from Kerr dynamics, but the ringdown has so far only been explored as an eigenvalue problem. We present the first time-domain implementation of this framework and perform (2+1)-dimensional scattering experiments with Gaussian wave packets. This approach provides the full linear evolution of the perturbation, from the initial prompt response through the ringdown and into the late-time regime. Using frequency-domain eigenvalue results as benchmarks, we find excellent agreement for low azimuthal numbers, while higher azimuthal numbers are affected by mode mixing, which limits the precision of the extracted modes. We further use the time-domain waveforms to estimate quadratic and mixed coefficients within the linearized modified-potential model, providing a diagnostic for the regime of validity of the linear approximation. Beyond mode frequencies, we show that the deformation parameters can strongly affect the ringdown amplitude and phase, with trends that can be understood from the near-horizon structure of the modified potential. We also analyze the late-time behavior, finding that near-horizon deformations leave the tail exponent unchanged but can substantially shift the onset of the power-law decay. These results demonstrate that time-domain evolutions provide a complementary and flexible framework for testing parametrized deviations from General Relativity in black-hole perturbation theory.
\end{abstract}

\maketitle

\section{Introduction}

Advances in gravitational-wave (GW) measurements from binary black hole (BH) mergers, exemplified by GW250114~\cite{LIGOScientific:2025rid,LIGOScientific:2025wao}, are transforming BH observations from pioneering detections into precision tests of strong-field gravity~\cite{Berti:2025hly}. A key pillar of this program is BH spectroscopy~\cite{Dreyer:2003bv}, which tests the standard assumption that astrophysical BHs are described by the Kerr solution~\cite{PhysRevLett.11.237}. 

The linear response of a BH to an external perturbation proceeds through three main stages. There is an initial burst that depends on the source~\cite{Vishveshwara:1970zz,Leaver:1986gd,Andersson:1996cm}, a ringdown dominated by a discrete spectrum of complex frequencies known as quasinormal modes (QNMs)~\cite{Kokkotas:1999bd,Berti:2009kk,Berti:2025hly}, and a late-time power-law tail governed by the long-range structure of the spacetime~\cite{Price:1971fb,Gundlach:1993tn}. For Kerr BHs, linear perturbations are governed by the Teukolsky equation~\cite{Teukolsky:1973ha}, whose QNM spectrum and late-time behavior are entirely determined by the BH parameters~\cite{Berti:2025hly}.

Theories beyond General Relativity (GR)~\cite{Moffat:2005si,Capozziello:2011et,Berti:2015itd,Barack:2018yly,Fernandes:2022zrq} typically modify the Einstein-Hilbert action by introducing additional degrees of freedom, associated for instance with scalar, vector, or tensor fields, or by adding higher-derivative curvature terms built from the Riemann tensor. Such modifications can change both the BH background and its perturbative dynamics. BHs may therefore acquire additional properties, either in the form of independent charges, or primary hair, or quantities determined by the usual Kerr parameters, often referred to as secondary hair~\cite{Coleman:1991ku,Sotiriou:2015pka,Herdeiro:2015waa}. These possibilities challenge the Kerr/no-hair paradigm~\cite{Israel:1967wq,Carter:1971zc,Hawking:1971vc,Robinson:1975bv} and can leave characteristic imprints on the different stages of the perturbative response~\cite{Cardoso:2009pk}. 

There are several complementary approaches to studying deviations from GR through BH perturbations~\cite{Franchini:2023eda}. In kinematical approaches, the deformation is introduced at the level of the background metric, and its impact on the BH response is probed using test fields, such as scalar~\cite{Volkel:2019muj,Pani:2026yzi} or electromagnetic perturbations. Closely related are eikonal methods~\cite{Dolan:2010wr,Konoplya:2017wot}, which connect QNMs to the properties of photon orbits without requiring the full perturbation equations of a specific theory. A second route is theory-specific and dynamical~\cite{Li:2022pcy,Cano:2023tmv}: the perturbation equations are derived directly from a given modified-gravity theory, thereby capturing additional degrees of freedom, modified couplings, and possible mixing between perturbation sectors. 
A third route is parametrized and dynamical. Deviations are introduced directly into effective perturbation equations, such as the Regge-Wheeler/Zerilli~\cite{Regge:1957td,Zerilli:1970wzz} or Teukolsky equation~\cite{Teukolsky:1973ha}, in a theory-agnostic manner~\cite{McManus:2019ulj,Cardoso:2019mqo,Franchini:2022axs,Cano:2024jkd}. This formulation can be benchmarked against theory-specific calculations while remaining sufficiently flexible to model broad classes of modifications, including environmental effects~\cite{Pezzella:2024tkf,Spieksma:2024voy}. It can also be studied in both the frequency domain and in the time-domain evolution~\cite{Thomopoulos:2025nuf,Nakashi:2025fbr}.

Modified Teukolsky-type equations in the frequency domain have been derived in many beyond-GR settings~\cite{Li:2022pcy,Hussain:2022ins,Cano:2023tmv}. These methods have subsequently been used to study isospectrality breaking~\cite{Li:2023ulk} and perturbations of slowly rotating black holes in dynamical Chern-Simons gravity~\cite{Wagle:2023fwl,Li:2025fci}. Building on related developments in higher-derivative gravity~\cite{Cano:2023tmv}, the parametrized QNM framework for modified Teukolsky equations~\cite{Cano:2024jkd} expresses a broad class of separable radial modifications as a power-series deformation of the Teukolsky potential. It provides the rotating-BH analog of parametrized QNM frameworks for Schwarzschild black holes~\cite{Cardoso:2019mqo,McManus:2019ulj,Volkel:2022aca,Volkel:2022khh,Hirano:2024fgp,Franchini:2022axs,Kimura:2020mrh}. Here, the modified Teukolsky potential is expressed in terms of powers of the radial Boyer-Lindquist coordinate. It is linear in a new set of parameters that depend on the specific theory beyond GR. 

Frequency-domain methods, formulated as boundary-value problems, efficiently determine QNM spectra and perturbative frequency shifts, but do not provide the complete evolution arising from prescribed initial data. A time-domain formulation instead gives simultaneous access to the prompt response, QNM ringing, ringdown amplitudes and phases, mode excitation and mixing, and the transition to the late-time tail~\cite{Vishveshwara:1970zz,Price:1971fb,Krivan:1997hc,Pazos-Avalos2004,Doneva:2020nbb,Pedrotti:2024znu,Thomopoulos:2025nuf,DeSimone:2026mkz,Silva:2024ffz,Silva:2026jih,Konoplya:2022pbc}. 

In this work, we present the first time-domain implementation of the parametrized modified Teukolsky equation introduced in Ref.~\cite{Cano:2024jkd}. We investigate its dynamics through scattering experiments with Gaussian wave packets, building on previous time-domain formulations of the standard Teukolsky equation~\cite{Krivan:1997hc,Pazos-Avalos2004}.

We extract the fundamental QNM frequencies from the numerical waveforms and compare them with the linear frequency-domain predictions~\cite{Cano:2024jkd} for small deformation parameters. We find excellent agreement for low azimuthal numbers, while the extraction of modes with larger $m$ is limited by spherical–spheroidal mode mixing. We then determine quadratic and mixed QNM corrections within the linearized modified-potential model. Although these terms do not capture the complete second-order dynamics of a generic modified-gravity theory, they provide a quantitative diagnostic of the parameter range over which the linear expansion remains reliable.

Finally, going beyond corrections to the QNM spectrum, time-domain simulations enable us to investigate the impact of the deformed potential on the amplitudes and phases of the perturbation and on the late-time tail. We relate the QNM amplitudes and phases to the deformation parameters of the modified Teukolsky potential, finding similarly strong modifications as in the non-rotating case~\cite{Thomopoulos:2025nuf}, and analyze the tail exponent for different classes of deformations, finding agreement with recent results in the literature~\cite{Rosato:2025rtr}.

Our work is organized as follows. In Sec.~\ref{sec: II}, we present the modified Teukolsky equation in both frequency and time domains. Sec.~\ref{sec: III} compares frequency- and time-domain QNM estimates at linear order in the deformation parameter, while Sec.~\ref{sec: IV} computes the second-order coefficients of the linear modified Teukolsky equation. In Sec.~\ref{sec: V}, we analyze the effects of the modified potential on amplitudes, phases, and late-time tails. We conclude in Sec.~\ref{Conclusions}. Appendix~\ref{Appendix_A} provides additional details on the numerical time-domain implementation and Appendix~\ref{Appendix_B} provides comparisons of retrograde QNMs. 
In all applications and figures, we report results in units of $M=0.5$ with angular momentum parameter $a \in [0,0.5]$. Throughout this paper, we adopt geometric units $G_N=c=1$.

\section{Modified Teukolsky equation}
\label{sec: II}

In this section, we describe the modified Teukolsky equation in the frequency- (Sec.~\ref{subsec: frequency domain}) and time-domain (Sec.~\ref{subsec: time domain}) formulations. 

\subsection{Frequency domain}
\label{subsec: frequency domain}

\begin{figure*}[ht]
    \centering
    \includegraphics[width=\linewidth]{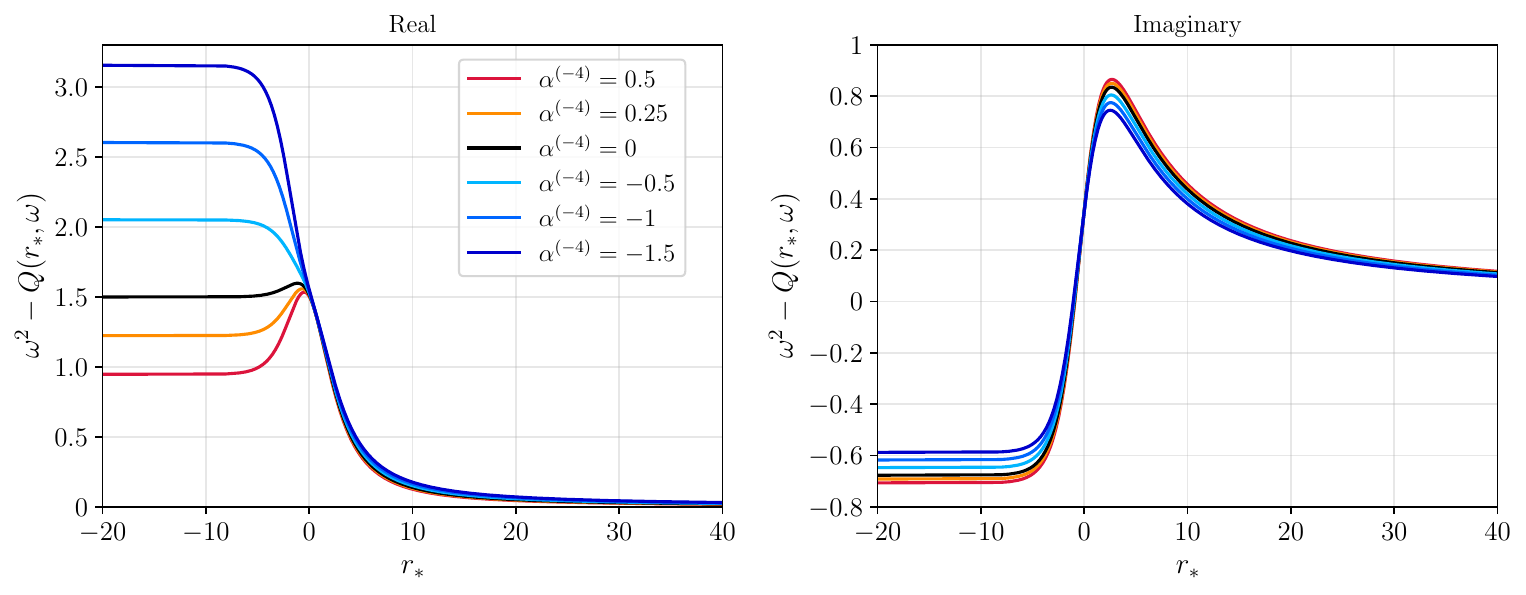}
    \caption{Real and imaginary parts of the modified Teukolsky potential for angular momentum $M=0.5,\,a=0.35, \,\ell=2,\,m=2,\,s=-2$, $k=-4$ and several values of $\alpha^{(k)}$, where $\omega$ corresponds to the modified fundamental QNM, respectively.}
    \label{fig:potential_l2m2_k-4_real}
\end{figure*}

Under the assumption of small deviations from GR, the modified radial Teukolsky equation for a spin $s$ field proposed in~\cite{Cano:2024jkd} has the general form
    \begin{equation}\label{Mod_Teu_equation}
        \frac{1}{\Delta^sR(r)}\frac{d}{dr}\left[\Delta^{s+1}\frac{dR}{dr}\right]+V(r)+\delta V(r)=0\,,
    \end{equation}
where $\Delta=r^2-2Mr+a^2$ depends on the mass $M$ and angular momentum $a$ of the BH, and $V(r)$ is the standard complex, frequency-dependent Teukolsky potential~\cite{Teukolsky1972} 
\begin{equation}
    V(r) = 2is\frac{dK(r)}{dr}-\lambda_{\ell m}+\frac{1}{\Delta}\left(K(r)^2-isK(r)\frac{d\Delta}{dr}\right)\,,
\end{equation}
where
\begin{align}
    K(r)&=(r^2+a^2)\,\omega-am\,,\\
    \lambda_{\ell m} &= B_{\ell m}+a^2\omega^2-2am\omega\,,
\end{align}
are functions of the frequency $\omega$ and the separation constant $B_{\ell m}$.
The modification in the radial potential is instead given by
\begin{equation}\label{Mod_potential}
    \delta V(r)=\frac{1}{\Delta}\sum_{k=-K}^{4}\alpha^{(k)}\left(\frac{r}{r_+}\right)^k\,,
\end{equation}
and depends on the position of the outer event horizon of the Kerr BH $r_+=M+\sqrt{M^2-a^2}$ and a set of coefficients that can in principle be complex $\alpha^{(k)}=\alpha_R^{(k)}+i\alpha^{(k)}_I$, while $K$ is a positive integer. This type of potential can be found in effective field theories (EFTs) and higher-derivative theories of gravity~\cite{Cano:2023tmv,Cano:2024ezp}. 

In terms of the tortoise coordinate
\begin{equation}\label{tortoise_coordinate}
    \frac{\dd r_*}{\dd r}=\frac{r^2+a^2}{\Delta}\,,
\end{equation}
the modified Teukolsky equation can be recast as~\cite{Tang:2025qaq}
\begin{equation}\label{ModTeu_tort}
    \frac{d^2Y}{d r_*^2}+\left[(V+\delta V)\frac{\Delta}{(r^2+a^2)^2}-G^2-\frac{dG}{dr_*}\right]Y=0\,,
\end{equation}
with 
\begin{align}
    Y&=\Delta^{s/2}(r^2+a^2)^{1/2}R(r),\\
    G&=\frac{s(r-M)}{r^2+a^2}+\frac{r\Delta}{(r^2+a^2)^2}\,.
\end{align}
The new potential
    \begin{equation}
        Q(r_*,\omega)=(V+\delta V)\frac{\Delta}{(r^2+a^2)^2}-G^2-\frac{dG}{dr_*}
    \end{equation}
shows a different behavior near the horizon ($r_*\to-\infty$) and at large distance ($r_*\to \infty$) depending on the index $k$. In particular, for $k<3$ the modification mainly affects the near-horizon behavior~\cite{Cano:2024bhh}, while the potential at large distance scales as $1/r$ as for the Kerr BH. On the other hand, $k=3$ introduces a new contribution that scales as $1/r$, so it can affect how the potential decays at infinity. Finally, $k=4$ adds a constant term to the potential at large distances.

The real and imaginary part of $\omega^2-Q(r_*,\omega)$ is shown in Fig.~\ref{fig:potential_l2m2_k-4_real} for selected values of $\alpha^{(-4)}$. The $\omega^2$ term is added in such a way that the potential approaches zero at large distances. The most interesting feature of the real part of the potential is the shift in the asymptotic value near the horizon. In particular, for sufficiently large negative values of $\alpha^{(-4)}$, the potential at the horizon dominates over the peak, while for sufficiently large and positive values of $\alpha^{(-4)}$, the value of the potential near the horizon decreases. In contrast, the imaginary part of the potential is not affected significantly by the deformation. The imaginary potential peak increases with $\alpha^{(k)}$, while the asymptotic value at the horizon exhibits the opposite behavior.

\subsection{Time domain}
\label{subsec: time domain}

\begin{figure*}[ht]
    \centering
    \includegraphics[width=\linewidth]{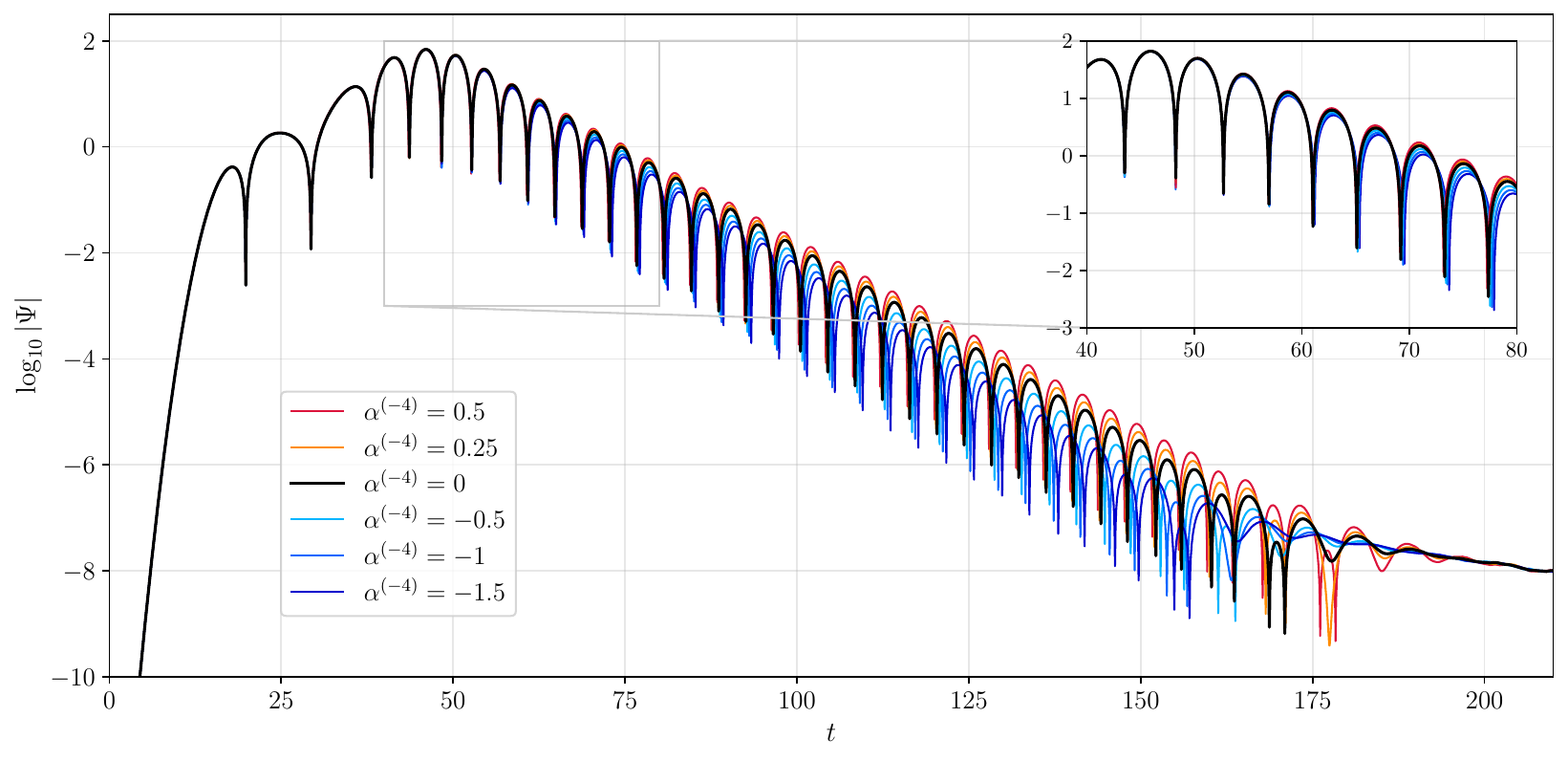}
    \caption{Waveforms obtained for $M=0.5,\,a=0.35,\,\ell=2,\,m=0,\,k=-4$ across several values of $\alpha^{(-4)}$. The inset shows the waveforms shortly after the peak.}
    \label{fig:waveforms}
\end{figure*}

The time-domain vacuum Teukolsky equation for a field $\psi(t,r,\theta,\phi)$ is~\cite{Teukolsky1972, Krivan:1997hc}
\begin{align}\label{master_equation_Kerr}
        &\left[\frac{(r^2+a^2)^2}{\Delta}-a^2\sin^2\theta\right]\frac{\partial^2\psi}{\partial t^2}+
        \left[\frac{a^2}{\Delta}-\frac{1}{\sin^2\theta}\right]\frac{\partial^2\psi}{\partial \phi^2}+ \nonumber\\& \frac{4Mar}{\Delta}\frac{\partial^2\psi}{\partial t\partial\phi}-2s\left[\frac{M(r^2-a^2)}{\Delta}-r-ia\cos\theta\right]\frac{\partial \psi}{\partial t}+
\nonumber\\&-\frac{1}{\sin\theta}\frac{\partial}{\partial \theta}\left(\sin\theta\frac{\partial\psi}{\partial\theta}\right)-2s\left[\frac{a(r-M)}{\Delta}+\frac{i\cos\theta}{\sin^2\theta}\right]\frac{\partial \psi}{\partial \phi}+\nonumber\\
        & -\Delta^{-s}\frac{\partial}{\partial r}\left(\Delta^{s+1}\frac{\partial \psi}{\partial r}\right)
        +(s^2\cot^2\theta-s)\psi=0\,,
\end{align}
where scalar, electromagnetic and gravitational perturbations correspond to $s=0,-1,-2$, respectively. In this paper, we are interested in gravitational perturbations. Eq.~\eqref{Mod_Teu_equation} and Eq.~\eqref{master_equation_Kerr} are related via the ansatz $\psi=e^{-i\omega t} e^{im\phi}S(\theta)R(r)$, exploiting the separability of the Teukolsky equation~\cite{Teukolsky1972}.

Due to the small coupling approximation, the deformation in Eq.~\eqref{Mod_Teu_equation} only affects the last term in Eq.~\eqref{master_equation_Kerr} 
\begin{equation}
    V_T=s^2\cot^2\theta-s\,.
\end{equation}
which depends explicitly on the spin weight. In the modified Teukolsky equation, the potential becomes 
\begin{equation}
    V'_T(r)=s^2\cot^2\theta-s-\delta V(r)
\end{equation}
as a function of the radial coordinate. 
It is worth mentioning that the deformed potential alters the discrete symmetries of the Teukolsky equation~\cite{DeSimone:2026xnz}, which can lead to the breaking of the $m=0$ degeneracy for complex-valued deformations $\alpha^{(k)}$.

The modified Teukolsky equation is solved numerically via the modified Lax-Wendroff method~\cite{Krivan:1997hc}, which is second-order convergent~\cite{Pazos-Avalos2004}. Details on the numerical implementation can be found in Appendix~\ref{Appendix_A}.

To obtain stable time evolutions, we adopt the tortoise coordinate $r_*$ given in Eq.~\eqref{tortoise_coordinate} and the Kerr azimuthal coordinate $\tilde\phi$
    \begin{equation}\label{Kerr_coord}
        \dd\tilde\phi=\dd\phi+\frac{a}{\Delta}\dd r\,,
    \end{equation}
which removes the pathological behavior of the Boyer-Lindquist coordinates near the event horizon~\cite{Krivan:1996da}.
Furthermore, given the axisymmetric nature of the spacetime, the field can be decomposed as 
    \begin{equation}\label{angular_decomposition}
        \psi(t,r_*,\theta,\tilde\phi) = \Psi(t,r_*,\theta)\,e^{im\tilde\phi},
    \end{equation} 
making the integration problem (2+1)-dimensional. Eq.~\eqref{master_equation_Kerr} is thus cast as a system of two coupled first-order differential equations for the field $\Psi$ and $\Pi$
\begin{equation}
    \Pi = \partial_t\Psi+b(r,\theta)\,\partial_{r_*}\Psi,
\end{equation}
where 
\begin{align}
    b(r, \theta) &= \frac{r^2+a^2}{\Sigma}\,,\\
    \Sigma^2 &= (r^2+a^2)^2-a^2\Delta\sin^2\theta\,.
\end{align}
The numerical integration has been performed over a two-dimensional grid $(r_*,\theta)$, using $N_r=8000$ points in the radial direction and $N_\theta=32$ points in the angular direction. The inner numerical boundary is placed at $r_*=-200M$, and the outer one is set to $r_*=400M$.

The boundary conditions in the tortoise coordinate have been imposed compatibly with a QNM type of problem, i.e., ingoing advection equation at the horizon and outgoing at large distance. As for the angular direction, we require that at the poles $\theta=0,\,\pi$ the field satisfies $\partial_\theta\Psi=0$ for $m$ even, and $\Psi=0$ for $m$ odd~\cite{Pazos-Avalos2004}. The initial data have been chosen as a combination of ingoing and outgoing Gaussian wave packet modulated by the $\theta$ component of the spin-weighted spherical harmonics
    \begin{align}
        \Psi|_{t=0}&=\sqrt{\frac{2\ell+1}{4\pi}}  d^{(s)}_{\ell m}(\theta) \,e^{-(r_*-r_*^0)^2/2}\,,\\
        \Pi|_{t=0}&=0\,,
    \end{align}
written in terms of the Wigner $d$-functions $d^{(s)}_{\ell m}(\theta)$~\cite{Campbell:1971rm}. Here, the Gaussian is centered around $r^0_*=12M$ while the signal is extracted at $r_*=30M,\theta=\pi/2$. Notice also that, since the Teukolsky equation is linear, the amplitude of the initial data does not affect the relative mode excitation~\cite{Kholyavka:2026uam}. 
For the purposes of time-domain simulations, the BH mass has been set to $M=0.5$ since it is a scaling parameter. 

It is important to highlight that the spin-weighted spherical harmonics do not represent the eigenfunctions of the angular Teukolsky equation (except for $a=0$). Those are instead given by the spin-weighted spheroidal harmonics~\cite{Berti:2005gp}, which depend on  $(\omega,s,a)$. As a consequence, the simulations are affected by the \textit{mode mixing} phenomenon~\cite{Krivan:1997hc, Hod:1999rx}, where additional modes are excited during the time evolution due to the choice of imperfect initial data and the rotation of the spacetime. 

From a physical perspective, this effect is due to the frame dragging of inertial frames ~\cite{Hod:1999rx}. Mathematically, instead, it can be ascribed to the fact that spheroidal harmonics depend on both angular momentum $a$ and angle $\theta$. Aside from the physical mode mixing, there is also a ``numerical mode mixing''~\cite{Pazos-Avalos2004} that should be taken into account, associated with the angular and radial resolution of the simulations. This is especially relevant for high multipoles $(\ell,m)$ where the angular functions exhibit many rapid oscillations in the $\theta$ direction. The important difference between the two types of mixing is that the numerical one becomes less prominent as resolution increases, while the physical one remains unaffected. 

To assess the effect of numerical and physical mode mixing, we have studied the impact of higher radial or angular resolution on the QNM extraction. In particular, simulations with twice the radial number of points $N_r=16000$ and more points in the angular direction $N_\theta=40$ lead to modest improvements in the QNM extraction of $\leq 0.3\%$ for the real part and negligible for the imaginary part. This result suggests that the effect of numerical mode mixing is subdominant even at the original resolution and that the error in the QNM estimates is mainly due to the physical mode mixing.

Several waveforms are shown in Fig.~\ref{fig:waveforms} for a single deformation $k=-4$ and many values of $\alpha^{(-4)}$. It is interesting to notice that, for this value of $k$, the most relevant deviations from GR appear in the QNM ringing phase. At the same time, early- and late-time evolutions are not significantly affected by the deformed potential. Similar profiles are observed for all values of $k<3$.

\section{Frequency vs time domain}
\label{sec: III}

For the purpose of comparing frequency- and time-domain QNM estimates, it is important to note that the modes obtained from time-domain simulations are affected by systematic errors arising from the numerical integration scheme (grid resolution, mode mixing, etc.). Non-QNM contributions and the choice of the fitting window can
introduce additional systematics~\cite{Baibhav:2017jhs,Bhagwat:2019dtm,Baibhav:2023clw,Nee:2023osy,Zhu:2023mzv,Giesler:2024hcr,Volkel:2025jdx}.

In the Kerr case, this error can be quantified exactly because the correct QNMs are known. Since mode mixing is minimal for $m=0$, the extracted Kerr QNMs agree with the eigenvalue predictions to $\sim0.3\%$. For larger $m$, mode mixing becomes more relevant with relative errors $\sim 5-6\%$ for the real part of the $\ell=m=2$ mode and $\sim 1\%$ for the imaginary part at $a=0.35$. Similar results are obtained for the retrograde modes $(m<0)$. 

The systematic errors are also affected by the deformation $\alpha^{(k)}$ in the potential. Since for a generic deformation of the Teukolsky potential, the exact QNMs are not known, it can be difficult to assess how the systematic error depends on the deformation. Nonetheless, for small values of $\alpha^{(k)}$, the systematic error does not change significantly, as the linear-order coefficients for low multipoles are always obtained with great accuracy. The effect can be more relevant for higher values of $\alpha^{(k)}$ and multipoles, since mode mixing will also lead to the excitation of additional modes.

Assuming small deviations from GR, the QNM frequencies and the angular separation constant for the modified Teukolsky equation can be expressed as~\cite{Cano:2024jkd}
    \begin{align}\label{Frequency_domain_QNMs}
        \omega_{n\ell m}&\simeq \omega^0_{n\ell m}+\sum_{k=-K}^4 d^{(k)}_{\omega,n\ell m}\alpha^{(k)}\,,\\
        B_{\ell m}&\simeq B_{\ell m}^0(a\omega)+\sum_{k=-K}^4 d^{(k)}_{B,\ell m}\alpha^{(k)}\,,
    \end{align}
in terms of the complex coefficients $d^{(k)}_{\omega,n\ell m}$ and $d^{(k)}_{B,\ell m}$, which have been computed in the frequency domain using Leaver's method and correspond to the leading-order Taylor expansion~\cite{Leaver:1985ax,Cano:2024jkd}. 
Note that the coefficient $d^{(k)}_{\omega,n\ell m}$ have units of $M^{-3}$ since $[\alpha^{(k)}]=M^{2}$ and in the following are expressed assuming $M=0.5$.
The numerical results of the framework are available in a \texttt{GitHub} repository~\cite{github} along with tutorials on how to apply it.

\begin{figure}[H]
    \centering
    \includegraphics[width=\linewidth]{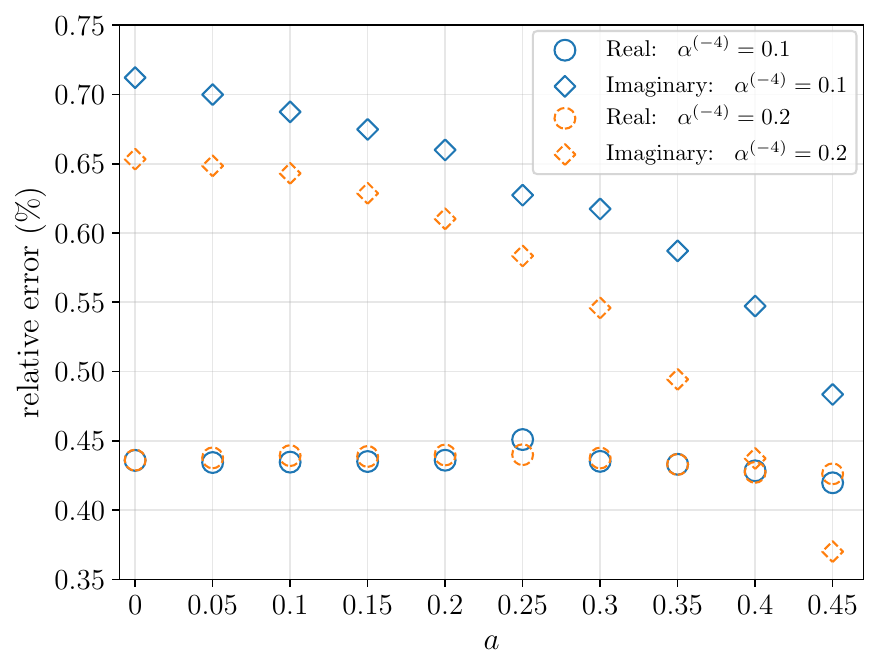}
    \caption{Relative error $(\%)$ of the real and imaginary parts of the QNMs obtained from frequency and time domain as a function of $a$ for $\ell=2,m=0,k=-4,\,\alpha^{(-4)}=0.1,0.2$.}
    \label{fig:QNM_comparison}
\end{figure}

The relative errors for the real and imaginary parts of the QNMs obtained from the Prony method and the linear-order frequency-domain estimates as a function of $a$ are shown in Fig.~\ref{fig:QNM_comparison}. Here, the relative error on the real part remains compatible with $\sim0.4\%$ for both values of $\alpha^{(-4)}$. In contrast, the relative error on the imaginary part decreases with $a$ and is consistently lower for $\alpha^{(-4)}=0.2$ than $\alpha^{(-4)}=0.1$.

\begin{figure*}[ht]
    \centering
    \includegraphics[width=\linewidth]{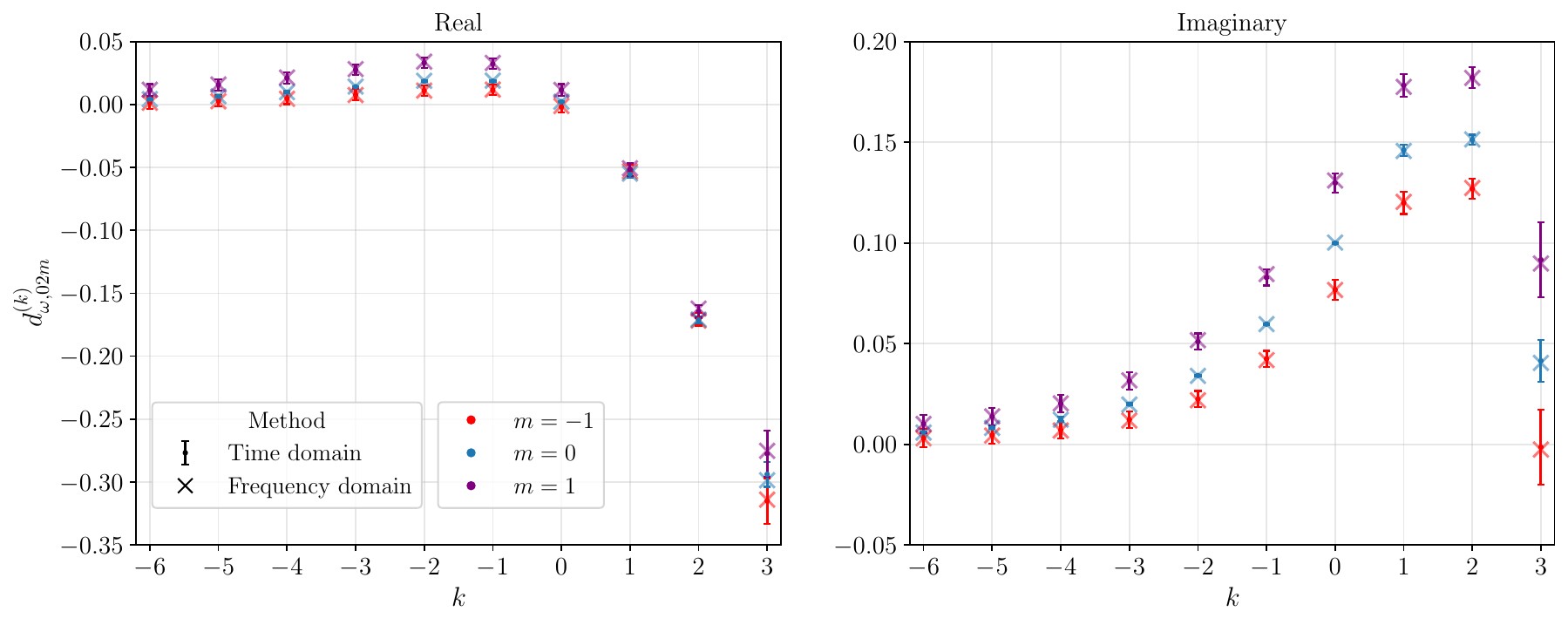}
    \caption{Comparison of the coefficients $d^{(k)}_{\omega,02m}$ for the real and imaginary parts of the QNMs from frequency and time domain at $a=0.25,\,\ell=2$ for $m=-1,0,1$. }
    \label{fig:d_l2m01}
\end{figure*}

\begin{figure*}[ht]
    \centering
    \includegraphics[width=\linewidth]{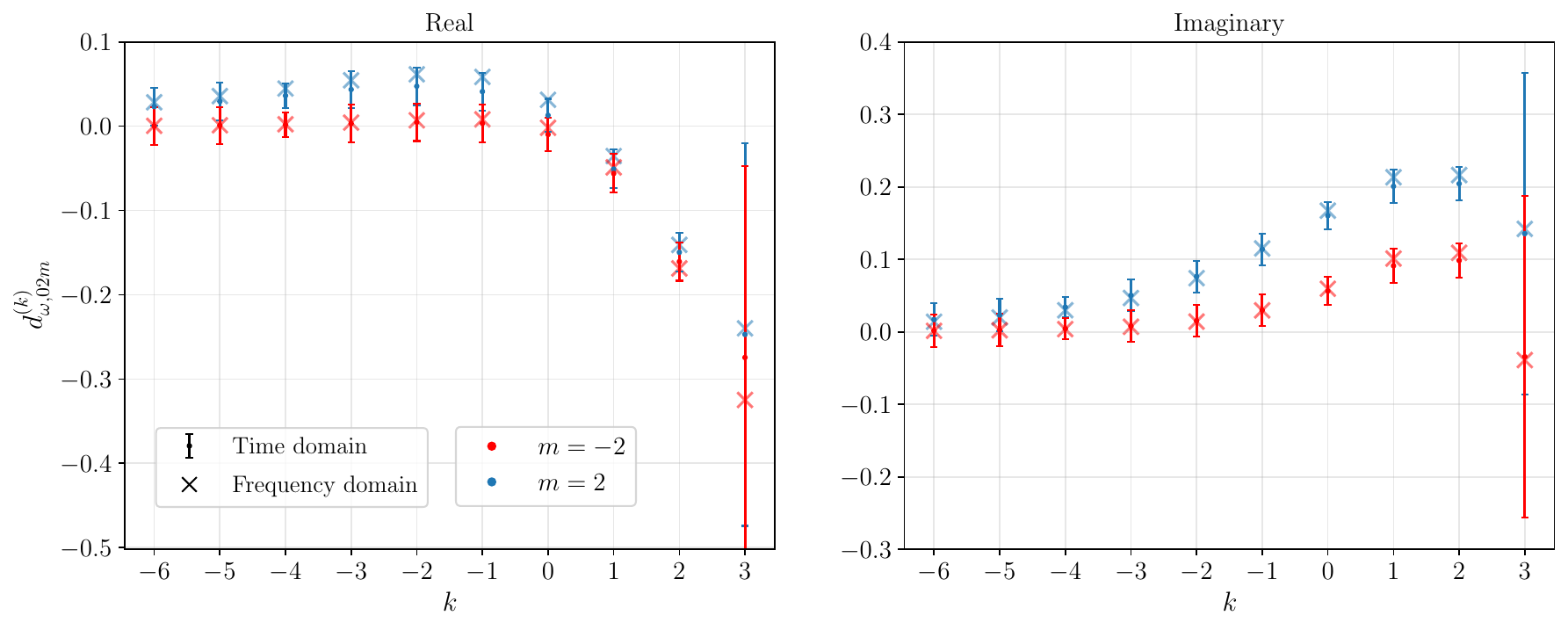}
    \caption{Comparison of the linear coefficients $d^{(k)}_{\omega,02m}$ for the real and imaginary parts of the QNMs from frequency and time domain at $a=0.25,\,\ell=2,\,m=-2,2$.}
    \label{fig:d_l2m2_a05}
\end{figure*}

The coefficients $d^{(k)}_{\omega,n\ell m}$ can also be estimated from the Modified Teukolsky equation in the time domain~\cite{Krivan:1997hc}. First, one has to fix the indices $(k,\ell,m)$ and run multiple simulations for selected values of $\alpha^{(k)}$ (for example five values of $\alpha^{(k)} \in [-0.1,0.1]$ around the GR case $\alpha^{(k)}=0$). Even though the $\alpha^{(k)}$ can be complex numbers, for the purposes of determining the linear $d^{(k)}_{\omega,n\ell m}$ coefficients, it is sufficient to consider real values of $\alpha^{(k)}$.

The QNMs are then extracted from the numerical waveforms using the Prony method~\cite{Berti2007}, i.e., by fitting the waveform with a sum of damped sinusoids. Here, we are mainly interested in the fundamental modes $(n=0)$ which can be reliably estimated from the waveforms. Once the QNMs are known, the linear coefficients $d^{(k)}_{\omega,n\ell m}$ can be determined via polynomial fits.

To estimate the error on the coefficients $d^{(k)}_{\omega,n\ell m}$, we have first determined the uncertainty on the QNM estimates by applying the Prony method for hundreds of starting times~\cite{DeSimone:2026mkz} during the ringing phase, choosing a time window such that the fit is not affected by the late-time tail. Then, we have computed the $68\%$ highest density intervals (HDIs)~\cite{DeSimone:2026mkz} and used those estimates in the polynomial fits. 

The best results are obtained for $\ell=2,m=0$, where prograde and retrograde modes coincide, and mode mixing effects are minimized. Figure~\ref{fig:d_l2m01} shows the comparison between the coefficients $d^{(k)}_{\omega,02m}$ from frequency and time domain for $a=0.25$ and $k\in[-6,3]$ over the range $m=-1,0,1$. The relative errors for both the real and imaginary parts are $\sim 2\%$. 

One can notice from Fig.~\ref{fig:d_l2m01} that the $m=-1,0,1$ coefficients follow qualitatively similar profiles in both the real and imaginary parts. As for the real part, the coefficients are more sensitive to $k$ in the range $k\in[0,3]$. The imaginary part, instead, tends to monotonically increase with $k$, except for $k=3$. Moreover, there is a distinguishable hierarchy among the three cases in terms of $m$ for both the real and imaginary parts.

\begin{figure}[ht]
    \centering
    \includegraphics[width=\linewidth]{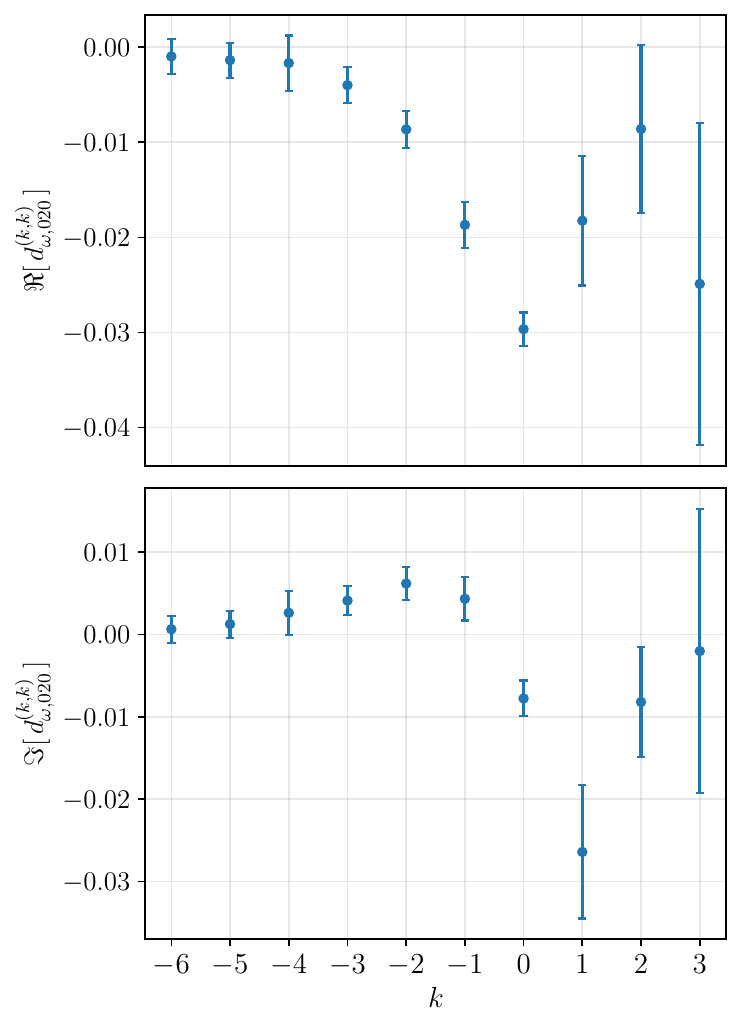}
    \caption{Real and imaginary part of the quadratic coefficients $d^{(k,\,k)}_{\omega,0\ell m}$ for $a=0.25,\,\ell=2,\,m=0$ and $k\in[-6,3]$.}
    \label{fig:d2_l2m0}
\end{figure}

Similar results have been obtained also for $a=0.25,\,\ell=2,\,m=\pm2$ in Fig.~\ref{fig:d_l2m2_a05} (the case $a=0.35$ is reported in Appendix~\ref{Appendix_B}). For these multipoles, the extracted QNMs have larger uncertainties due to mode mixing, and the $k=3$ estimates are not very accurate. In all the cases, the $k=4$ coefficients have not been computed because the simulations typically exhibit unstable behavior already at small values of $\alpha^{(k)}$. This can be ascribed to the large-distance nature of the deformation and the non-zero value of the potential asymptotically far from the BH.

\section{Quadratic coefficients}
\label{sec: IV}

The methodology developed in the previous section can also be used to compute the quadratic corrections to Eq.~\eqref{Frequency_domain_QNMs}. In particular, the equation can be generalized to higher orders as
\begin{align}\label{higher_order_ParTeu}
    \omega_{n\ell m}&\simeq \omega^0_{n\ell m}+\sum_k d^{(k)}_{\omega,n\ell m}\alpha^{(k)}+\frac{1}{2}\sum_{k_1} \sum_{k_2}d^{(k_1,\,k_2)}_{\omega,n\ell m}\alpha^{(k_1)}\alpha^{(k_2)}\nonumber\\&+\mathcal{O}(\alpha^{(k)\,3})\,, 
\end{align} 
where $d^{(k_1,\,k_2)}_{\omega,n\ell m}$ labels the second-order coefficients and the sum extends over the range $k\in[-K,4]$ as in Eq.~\eqref{Frequency_domain_QNMs}. Moreover, the coefficients $d^{(k_1,\,k_2)}_{\omega,n\ell m}$ are invariant under the exchange of $k_1$ and $k_2$ (i.e., $d^{(k_1,\,k_2)}_{\omega,n\ell m}$ = $d^{(k_2,\,k_1)}_{\omega,n\ell m}$) and the factor $1/2$ in the last term is added to avoid double counting the coefficients. Notice also that the second order coefficients have units $[d^{(k_1,\,k_2)}_{\omega,n\ell m}]=M^{-5}$ and they also have been computed assuming $M=0.5$. 

From a numerical standpoint, in the case of two deformations $\alpha^{(k_1)}$ and $\alpha^{(k_2)}$, the coefficients in Eq.~\eqref{higher_order_ParTeu} can be computed by creating a suitable grid of values for $\alpha^{(k_1)}$ and $\alpha^{(k_2)}$ around $\alpha^{(k_1)}=\alpha^{(k_2)}=0$, and running simulations for each of those values. The coefficients are then extracted by performing polynomial multidimensional fits of the QNM frequencies obtained for all the values of $\alpha^{(k_1)}$ and $\alpha^{(k_2)}$.

It should be noted, however, that the parametrized quasinormal mode framework for modified Teukolsky~\cite{Cano:2024jkd} relies on the assumption that the deformation in the potential is linear in $\alpha^{(k)}$. In fact, this is the type of equation found in many alternatives to GR for small values of the perturbative parameter~\cite{Li:2022pcy,Hussain:2022ins,Cano:2023tmv}. At higher order in $\alpha^{(k)}$, the modified Teukolsky equation does not have the same structure as in Eq.~\eqref{Mod_Teu_equation}: the radial and angular derivatives will have a different structure, perturbations of different spin-weights might be mixed, and new degrees of freedom might become relevant.

It is thus crucial to note that the quadratic coefficients are computed in the linear modified Teukolsky equation. The correct quadratic coefficients will, in general, contain additional contributions associated with the second-order modified Teukolsky equation, which can, in principle, be of comparable order of magnitude or larger.

Nonetheless, the quadratic coefficients computed within a linear model can still provide valuable information on the regime of validity of the linear approximation. In particular, the linear approximation will be valid as long as the linear contribution dominates over the second-order one. For each value of $k$, the quadratic coefficients can thus be used to determine the maximum value of $\alpha^{(k)}$ such that the linear approximation retains its validity. For a single non-zero $\alpha^{(k)}$, the comparison between first and second order contributions in Eq.~\eqref{higher_order_ParTeu} provides the following criterion for the validity of the linear approximation
    \begin{equation}\label{linear_validity}
        |d^{(k)}_{\omega,n\ell m}|>|d^{(k,\,k)}_{\omega,n\ell m}\alpha^{(k)}|\,,
    \end{equation}
written in terms of the absolute values since the coefficients $d^{(k)}_{\omega,n\ell m}$ are complex. If more than one $\alpha^{(k)}$ term is non-zero, then Eq.~\eqref{linear_validity} has to be modified and will, in general, include additional terms associated with the second-order coefficients.

Notice, however, that Eq.~\eqref{Mod_Teu_equation} can also be considered as a specific deformation of the Teukolsky equation without any direct connection to EFTs. From this perspective, it constitutes a spectral problem with different properties, similar to other studies in the literature where the Regge-Wheeler/Zerilli potential is modified by the addition of bumps in the potential~\cite{Berti:2022xfj,Cheung:2021bol} or other ad hoc deformations. It is thus reasonable to investigate the exact QNMs of the modified Teukolsky equation as well as the higher-order contributions in $\alpha^{(k)}$. Moreover, such operators can exhibit non-trivial features in the QNM spectrum and the late-time tail behavior, which are interesting from a phenomenological standpoint. Mass-like or asymptotically confining modifications can also support quasi-bound states and, in rotating spacetimes, superradiant
instabilities~\cite{Dolan:2007mj}. Thus, deformations of the Teukolsky potential can also give rise to quasi-bound states, and in some cases to instabilities, which could in principle be used to constrain viable parameters $\alpha^{(k)}$.  

\begin{figure}[ht]
\includegraphics[width=\linewidth]{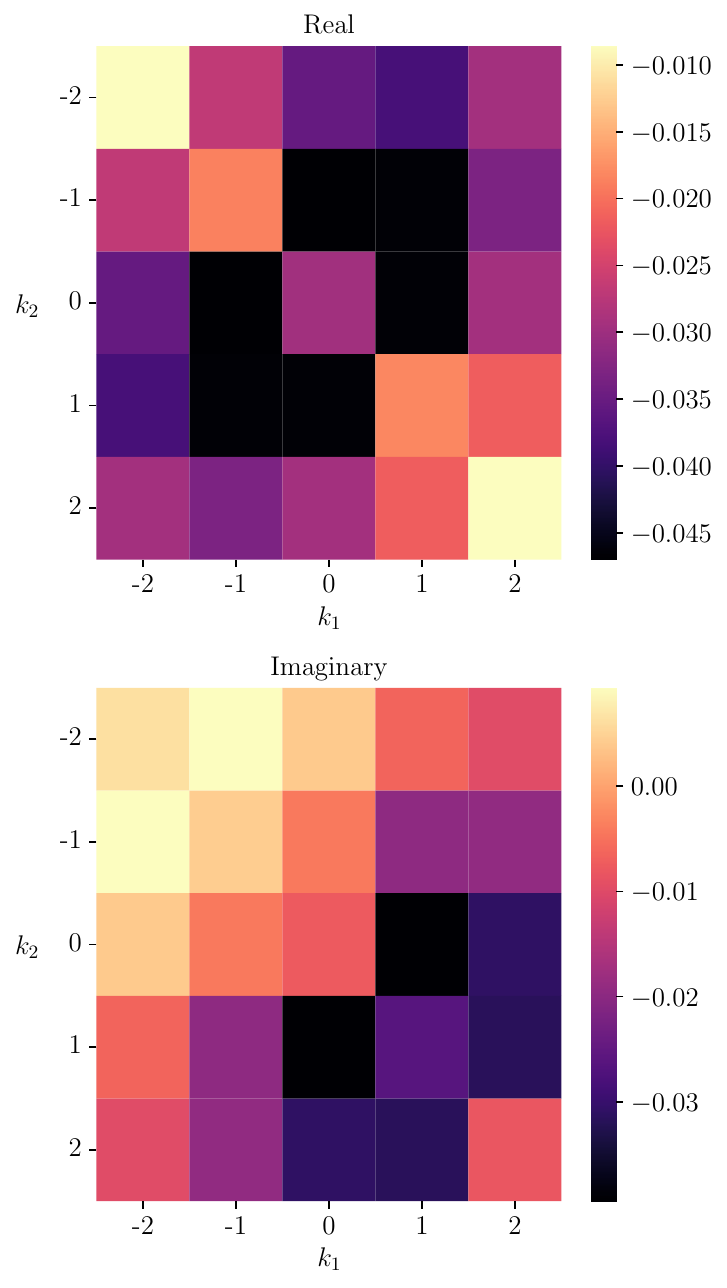}
\centering
\caption{Real and imaginary part of the mixed coefficients $d^{(k_1,\,k_2)}_{\omega,n\ell m}$ computed for $a=0.25,\,n=0,\,\ell=2,\,m=0$ in the range $k_1,\,k_2\in[-2,2]$ using the time-domain approach.}
\label{fig:heatmap_mixed_coef}
\end{figure}

The quadratic coefficients for $a=0.25,\,\ell=2,\,m=0$ are reported in Fig.~\ref{fig:d2_l2m0}. It is interesting to note that the error bars become larger for positive values of $k$, where the extraction of the second-order coefficients is less accurate. Similarly, the mixed coefficients $d^{(k_1,\,k_2)}_{\omega,n\ell m}$ for the fundamental mode $a=0.25,\,\ell=2,\,m=0$ are shown in Fig.~\ref{fig:heatmap_mixed_coef} for many values of $k_1$ and $k_2$. This figure provides a clear picture of how the coefficients change across $k_1$ and $k_2$ and their relative values. 

From the heat maps in Fig.~\ref{fig:heatmap_mixed_coef}, one can see that the mixed coefficients for the real and the imaginary parts are quite similar. Notice also that the quadratic fits provide additional estimates for the linear and quadratic coefficients for the cases $k_1=k_2$. For consistency, we have verified that they are always in agreement with the results obtained in the previous section. Additional details on the comparison between linear and quadratic coefficients for the mixed case can be found in Appendix~\ref{Appendix_B}.

\section{Additional properties of time-domain perturbations}
\label{sec: V}

In this section, we discuss the effects of the modified potential on the amplitude and phase of the perturbation field extracted using the Prony method (Sec.~\ref{subsec: amplitudes and phases}). Furthermore, in Sec.~\ref{subsec: tails} we study the late-time power-law tails of the modified Teukolsky potential.

\subsection{Amplitude and phase}
\label{subsec: amplitudes and phases}

Another advantage of the time-domain approach is the possibility to investigate how the ringdown amplitude and phase are affected by modifications to the Teukolsky potential. Both quantities can be estimated by applying the Prony method to the waveform extracted at a fixed distance from the BH. Moreover, in all simulations, we have used ingoing Gaussian initial data with fixed amplitude and width to study the effect of the deformed potential on them.

It is important to notice that, while the QNM spectrum of a Kerr BH is fully characterized by mass and angular momentum, in BH mergers the amplitude and phase of the field depend on the properties of the progenitors~\cite{Kamaretsos:2012bs}. Nonetheless, Gaussian initial data provide a controlled environment for studying the impact of the deformed potential on the field's amplitude and phase.

\begin{figure}[ht]
\includegraphics[width=\linewidth]{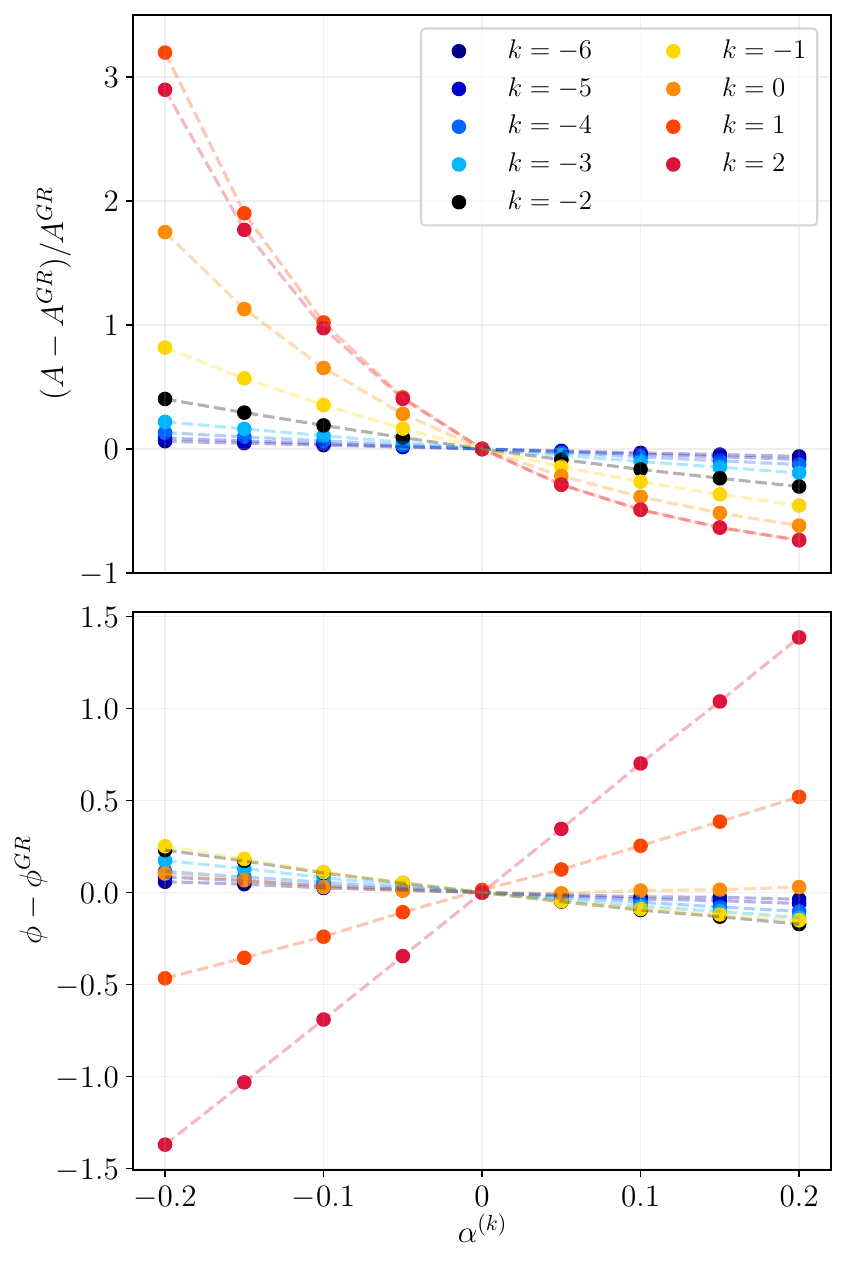}
\centering
\caption{Fractional deviation of the amplitude $A$ (top panel) and phase difference (bottom panel) of the modified Teukolsky equation compared to GR as a function of $\alpha^{(k)}$ for $k\in[-6,2]$ for $a=0.25,\,\ell=2,\,m=0$.
}
\label{fig:amp_ph_ModTeu}
\end{figure}

In Fig.~\ref{fig:amp_ph_ModTeu}, we show the relative error on the amplitude $A$ extracted from the waveform and the difference in phase $\phi$ compared to the GR value, for $k\in [-6,2]$. Larger values of $k$ are not reported since they give rise to instabilities over the range of $\alpha^{(k)}$ values considered. The amplitudes display a nonlinear profile for larger $k$ even at low values of $\alpha^{(k)}$, where also the deviations from GR are more pronounced.

The amplitude is strongly suppressed for positive values of $\alpha^{(k)}$, while it increases by a similar factor for negative values. This result can be explained by considering the behavior of the modified Teukolsky potential as a function of $\alpha^{(k)}$ in Fig~\ref{fig:potential_l2m2_k-4_real}. From a scattering perspective, as the potential plateau in the near-horizon region tends to increase for negative values, the transmitted component of the incoming wave will be suppressed. This leads to an enhancement of the reflected wave extracted from numerical simulations. In contrast, positive values of $\alpha^{(k)}$ will lower the potential plateau in the near-horizon region, thus enhancing the transmitted component and suppressing the reflected one. Positive values of $\alpha^{(k)}$ will thus be associated with smaller amplitudes compared to GR, and vice versa.

The phase shows a significantly different behavior. It grows with $\alpha^{(k)}$ for $k>0$ and does not exhibit significant departures from the linear profile across the considered range of $\alpha^{(k)}$.

\subsection{Tails}
\label{subsec: tails}

The last stage of the numerical time evolutions is dominated by a power-law profile~\cite{Gundlach:1993tp, Casals:2016soq, Gleiser:2007ti}, which can also be affected by the deformation in the Teukolsky potential. It is important to notice, however, that the tail profile depends on the behavior of the potential at large distances. Since, for $k<3$, the potential is mainly affected in the near-horizon region, the power-law tail exponent shows no significant departures from the Kerr case. This result is compatible with Ref.~\cite{Rosato:2025rtr}, where similar deformations in the Teukolsky potential have been considered. 

\begin{figure*}[ht]
\includegraphics[width=\linewidth]{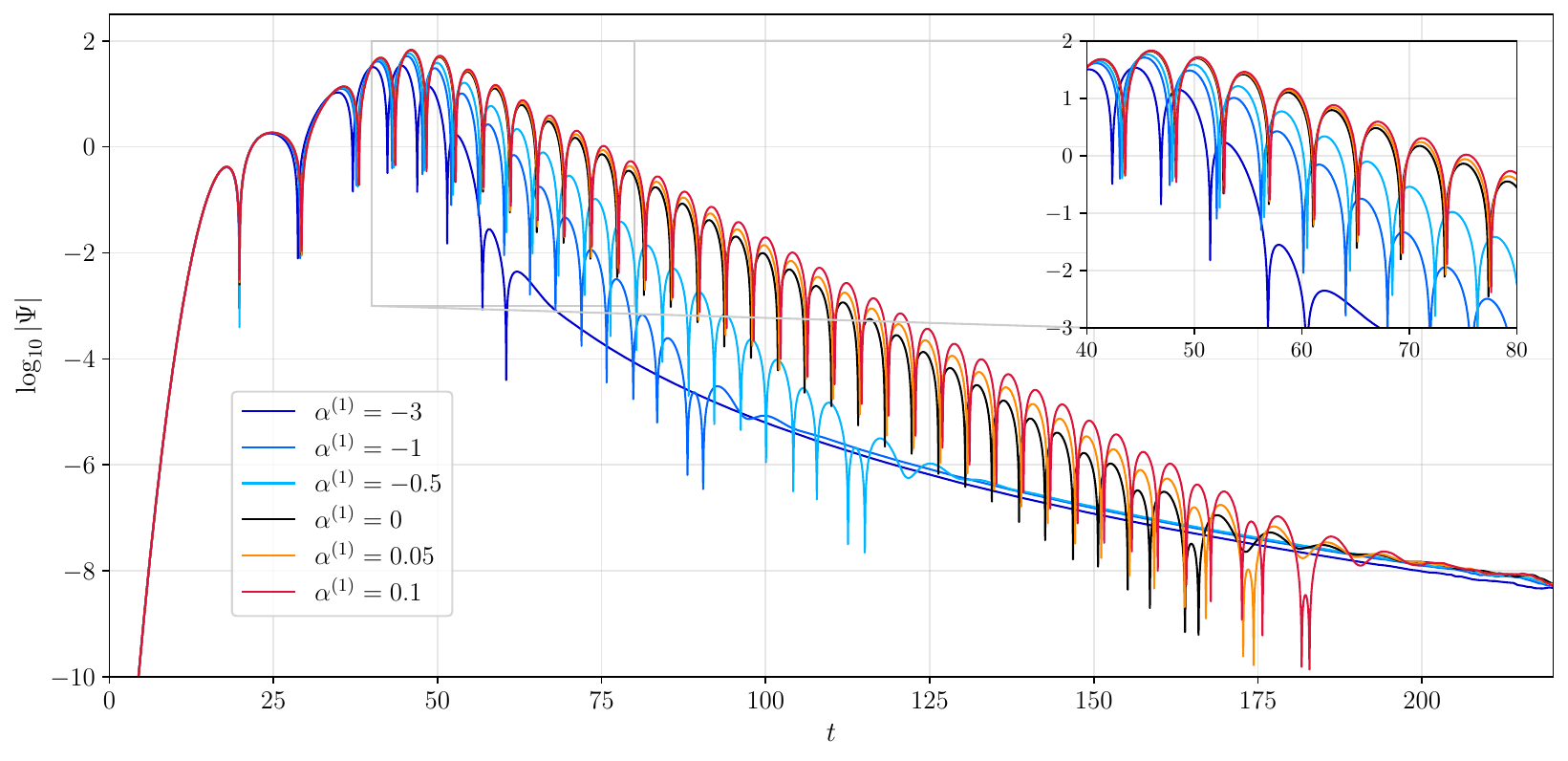}
\centering
\caption{Waveforms for $M=0.5,\,a=0.25,\,\ell=2,\,m=0, k=1$ for multiple values of $\alpha^{(1)}$. The inset displays the behavior of the waveforms near the peak.}
\label{fig:tail_k1}
\end{figure*}

Even if the power-law exponent is not affected by the deformed potential for $k<3$, one can notice from Fig.~\ref{fig:tail_k1} that the time onset of the tail can change significantly, depending on the value of $k$ and the deformation $\alpha^{(k)}$. In particular, for negative values of $\alpha^{(1)}$, the tail dominates at earlier times, while positive values of $\alpha^{(1)}$ delay the appearance of the power-law profile.

In Fig.~\ref{fig:tail_k1}, the case $\alpha^{(1)}=-3$ is used to show the effect of large deformations on the tail behavior, even though this value is typically outside the regime of validity of the model for theory-specific applications. It is interesting to note, however, that such large deformations also alter the early-time evolution of the field near the peak. 

This behavior is a consequence of how the QNMs are affected by the deformed Teukolsky potential. In particular, the damping time decreases for negative values of $\alpha^{(1)}$, corresponding to modes with shorter lifetime. In contrast, positive values of $\alpha^{(1)}$ lead to an increase in the damping time of the fundamental mode and thus to a longer ringing phase, while the effect on the mode frequency is subdominant.

It is worth stressing that, even if the code is not very accurate at late times since numerical noise becomes more relevant (and higher resolution is required), it is still possible to compare the tail profiles for the modified Teukolsky potential with the GR case. The behavior of the tail appears compatible with the GR tail scaling $\sim t^{-(2\ell+3)}$~\cite{Price:1971fb}, considering the $\sim10\%$ uncertainty in the tail extraction~\cite{Krivan:1997hc}.

\section{Discussion and Conclusions}
\label{Conclusions}

Time-domain methods are essential for properly characterizing the perturbative response of BHs and compact objects beyond GR. In this paper, we have presented the first time-domain implementation of the parametrized modified Teukolsky equation. The resulting framework provides a flexible and theory-agnostic approach for modeling deformations of the Teukolsky potential and quantifying their impact on the linear evolution of perturbations. 

As a demonstration, we studied the response of modified BHs to external perturbations via scattering experiments with Gaussian wave packets. This has enabled us to examine the impact of the modified Teukolsky potential on the propagation of initial data, the QNM ringing, and the late-time tails. In contrast to frequency-domain calculations, which efficiently determine the spectral properties of the modified operator, the time-domain approach gives direct access to the evolution generated by prescribed initial data. This makes it possible to study not only the QNM frequencies but also mode excitation and mixing, ringdown amplitudes and phases, and the transition from exponentially damped ringing to a power-law tail.

In the early part of the extracted waveform, we have found that the prompt response is quite robust to small deformations in the potentials. This is consistent with the near-horizon or large-distance character of the modifications of the potential. For sufficiently large deformations, however, visible changes can also arise near the peak of the waveform. The fundamental QNM frequencies, on the other hand, can exhibit substantial departures from the Kerr spectrum. First, we have estimated the coefficients of the linear QNM model for different multipoles. The comparison with the frequency-domain estimates shows excellent agreement for low azimuthal numbers $m$ and across a wide range of angular momenta. These comparisons provide an independent validation of the parametrized frequency-domain computations~\cite{Cano:2024jkd}, and clarify the range of multipoles for which direct QNM extraction from the present simulations is reliable.

We then used the time-domain waveforms to determine quadratic and mixed corrections to the QNM frequencies. These coefficients quantify the nonlinear dependence of the spectrum on the deformation parameters within the linearly modified Teukolsky equation and provide a practical diagnostic of the parameter range over which the linear QNM expansion remains accurate. They should not, however, be interpreted as the complete second-order corrections of a generic theory beyond GR. At higher orders, additional modifications of the perturbation operator, couplings between different perturbation sectors, and new dynamical degrees of freedom may contribute at comparable or larger magnitudes.

Beyond the QNM spectrum, we have also extracted the ringdown amplitudes and phases. These quantities are highly sensitive to beyond-GR effects, and their behavior has been linked to the near-horizon behavior of the modified potential. Moreover, we have also investigated the impact of the modified potential on the late-time tail, finding good agreement with existing literature and correlating the time onset of the tail with the QNM properties. The time when the tail becomes dominant can shift substantially because the deformation changes the damping rate and duration of the preceding QNM ringing. 

Overall, our results demonstrate that time-domain scattering experiments provide a powerful framework for studying modified Teukolsky equations and their associated beyond-GR signatures. These methods offer a complementary perspective to frequency-domain techniques and can help characterize the full linear response of modified BHs. It is worth noting that deformations of the Teukolsky operator can also lead to additional spectral properties, such as time-domain instabilities. These are primarily related to the behavior of the potential near the horizon and at large distances, and will be the focus of a forthcoming publication~\cite{Instabilities}. 

The approach presented in this paper can be generalized in many directions. First, it would be interesting to apply this framework to theory-specific models, i.e., higher-derivative gravity~\cite{Cano:2024ezp} or dynamical Chern-Simons gravity~\cite{Wagle:2023fwl,Li:2025fci}. More general deformations of the Teukolsky equation might also include non-separable terms in the angular component~\cite{Ghosh:2023etd}, frequency-dependent contributions, couplings to other fields, and multipole-dependent deformations~\cite{Tattersall:2017erk, Silva:2019scu}. Furthermore, higher-order modified Teukolsky equations could potentially be incorporated into this approach. Time-domain methods could also be employed to study overtones~\cite{Kubota:2025hjk} and explore other perturbative properties of BHs such as superradiance~\cite{Krivan:1997hc,Brito:2015oca}. From a numerical perspective, implementing more efficient integration techniques, such as those using hyperboloidal compactified coordinates~\cite{Harms:2013ib,Zhu:2023mzv}, would be of interest as well. This would potentially allow for stable evolutions over significantly longer time intervals, which are crucial for studying the impact of deformations on late-time tails and for investigations involving matter around the BH~\cite{Zenginoglu:2011zz,Burko:2016sfi,Laeuger:2025zgb,Steppohn:2025kbh}.

\section*{Acknowledgments}

C.D.S. and S.H.V. thank Pablo Cano for providing valuable feedback on the manuscript. C.D.S. is grateful to the University of T\"ubingen for hospitality during the realization of this work. C.D.S. and S.C. acknowledge the support of INFN, {\it sez. di Napoli}, {\it iniziative specifiche}  QGSKY and MOONLIGHT-2. K.D.K. and S.H.V. acknowledge funding from the European Union’s Horizon MSCA-2022 research and innovation programme
“EinsteinWaves” under grant agreement No. 101131233. The authors acknowledge support by the High Performance and Cloud Computing Group at the Zentrum für Datenverarbeitung of the University of Tübingen, the state of Baden-Württemberg through bwHPC and the German Research Foundation (DFG) through grant no INST 37/935-1 FUGG.

\bibliography{literature}

\appendix 

\section{Numerical implementation of the modified Teukolsky equation}
\label{Appendix_A}

In order to implement the modified Lax-Wendroff method~\cite{Krivan:1997hc}, the perturbation equation~\eqref{master_equation_Kerr} is written as a first-order system of partial differential equations
\begin{equation}\label{first_LW}
    \partial_t \textbf{u}+\textbf{M}\,\partial_{r_*}\textbf{u}+\textbf{L}\textbf{u}+\textbf{A}\textbf{u}=0,
\end{equation}
where the vector $\textbf{u}=\{\Psi_R,\Psi_I,\Pi_R,\Pi_I\}$ collects the real and imaginary parts of the fields $\Psi$ and $\Pi$. Moreover, the matrices $\textbf{M},\,\textbf{L},\,\textbf{A}$ are given by
\begin{equation}\label{LWmatrix1}
    \textbf{M}=\begin{pmatrix}
        b(r,\theta) & 0 & 0 & 0 \\
        0 & b(r,\theta) & 0 & 0 \\
        m_{31} & m_{32} & -b(r,\theta) & 0 \\
        -m_{32} & m_{31} & 0 & -b(r,\theta) \\
    \end{pmatrix},
\end{equation}
\begin{equation} \label{LWmatrix2}
    \textbf{A}=\begin{pmatrix}
        0 & 0 & -1 & 0 \\
        0 & 0 & 0 & -1 \\
        a_{31} & a_{32} & a_{33} & a_{34} \\
        -a_{32} & a_{31} & -a_{34} & a_{33} \\
    \end{pmatrix},
\end{equation}
\begin{equation}\label{LWmatrix3}
    \textbf{L}=\begin{pmatrix}
        0 & 0 & 0 & 0 \\
        0 & 0 & 0 & 0 \\
        l_{31} & 0 & 0 & 0 \\
        0 & l_{31} & 0 & 0 \\
    \end{pmatrix}.
\end{equation}
The coefficients in Eqs.~\eqref{LWmatrix1}-\eqref{LWmatrix3} are
\begin{align}
        m_{31} &= -b(r,\theta)\,c_1+b(r,\theta)\frac{\partial b}{\partial r_*}+c_2,\\ m_{32} &= b(r,\theta)\,c_3 - c_4,\\
        a_{33}&=c_1,\\ a_{34} &= -c_3,\\
        l_{31} &= c_7 \frac{\partial^2}{\partial \theta^2}+c_8\frac{\partial}{\partial \theta},
        \\a_{31}&=-c_5-\frac{\Delta}{\Sigma^2} \Re(\delta V),\label{LWdef1}
        \\ a_{32}&=-c_6+\frac{\Delta}{\Sigma^2}\Im(\delta V) \label{LWdef2},
    \end{align}
where the derivatives with respect to the coordinate $\theta$ are included in the coefficient $l_{31}$ and
    \begin{align}
        c_1 &=2s\frac{-3Mr^2+Ma^2+r^3+ra^2}{\Sigma^2},\\
        c_2 &=-2\frac{r\Delta(1+s)-M(a^2-r^2)s}{\Sigma^2},\\
        c_3 &=2a\frac{2Mrm+\Delta s\cos{\theta}}{\Sigma^2},\\
        c_4 &=-2am\frac{r^2+a^2}{\Sigma^2},\\
        c_5 &=\frac{\Delta (-m^2-2sm\cos{\theta}-s^2\cos^2{\theta}+s \sin^2{\theta})}{\Sigma^2 \sin^2\theta},\\
        c_6 &= -4\frac{(r-M)sma}{\Sigma^2},\\
        c_7 &= -\frac{\Delta}{\Sigma^2},\\
        c_8 &=-\cot{\theta}\frac{\Delta}{\Sigma^2}.
    \end{align}
The deformation of the Teukolsky potential affects only the coefficients in Eq.~\eqref{LWdef1} and Eq.~\eqref{LWdef2}, which are connected to terms in the Teukolsky equation that only depend on the field $\Psi$. Moreover, since the matrix $\textbf{M}$ is not affected by the deformation in the potential, the system of equations preserves its hyperbolic nature~\cite{Krivan:1997hc}.

Further details on the implementation of the Lax-Wendroff method can be found in Ref.~\cite{Krivan:1997hc}. Compared to Refs.~\cite{Krivan:1997hc, Pazos-Avalos2004}, the coefficients $a_{34}$ and $a_{31}$ have been corrected to account for some typos.

It is important to note that the behavior of the field near the horizon and at large distances is, in general, different from GR. As reported in~\cite{Cano:2024jkd}, the near-horizon field for the modified Teukolsky equation can be expressed as  
    \begin{equation}
        R(r)\sim\Delta^{-s}\,(r-r_+)^{-i\sigma},
    \end{equation}
where
    \begin{equation}
        \sigma=\frac{is}{2}+\sqrt{\left(\sigma_{GR}-\frac{is}{2}\right)^2+\frac{1}{4M^2-4a^2}\sum_{k=-K}^4\alpha^{(k)}},
    \end{equation}
    \begin{equation}
        \sigma_{GR}=\frac{2M r_+(\omega-m\,\Omega_H)}{(r_+-r_-)},\quad\quad \Omega_H=\frac{a}{2Mr_+}.
    \end{equation}
Here, $r_-=M-\sqrt{M^2-a^2}$ corresponds to the inner event horizon of the Kerr BH and $\Omega_H$ to the orbital frequency of the outer event horizon. This implies that the leading order behavior of the field near the horizon is still dictated by the term $\Delta^{-s}$, which is regular at the horizon. In contrast, the coefficients $\alpha^{(k)}$ enter the expressions for the propagating part of the field.

Even though the large-distance behavior of the field is also affected by the deformation, in the numerical implementation one can also adopt the approach of Ref.~\cite{Krivan:1997hc}. 
In fact, our computational domain is large enough that, for the time intervals associated with QNM ringing and late-time tails, the numerical evolutions are not affected by the effect of the boundary conditions.

\section{Complementary results}
\label{Appendix_B}

In this appendix, we first present additional comparisons between frequency-domain and time-domain estimates of the coefficients $d^{(k)}_{\omega,n\ell m}$. Then, we study in more detail the relationship between linear and quadratic coefficients when multiple deformation coefficients, $\alpha^{(k)}$, are present.

\begin{figure*}[ht]
\includegraphics[width=\linewidth]{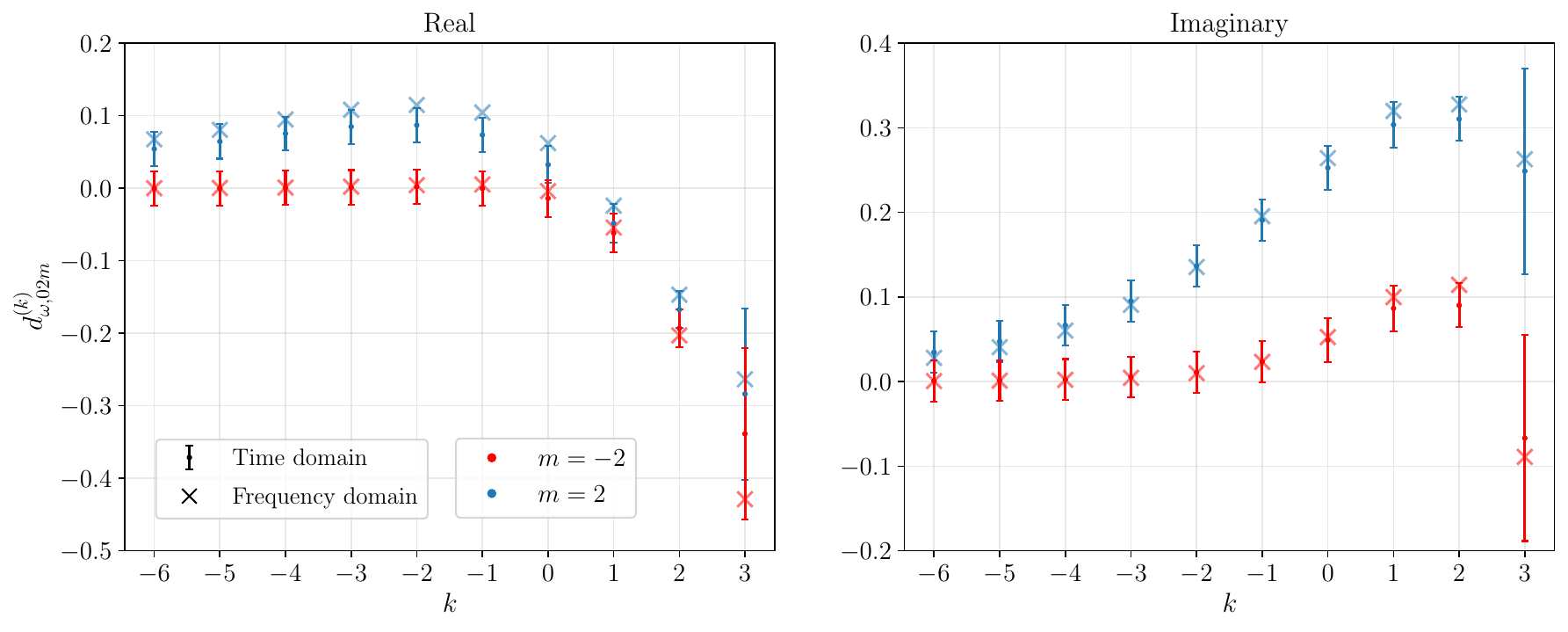}
\centering
\caption{Comparison of the linear coefficients from frequency and time domain at $a=0.35,\,\ell=2,\,m=-2,2$.}
\label{fig:d_l2m2_a07}
\end{figure*}

Figure~\ref{fig:d_l2m2_a07} shows the linear coefficients for prograde and retrograde modes for $a=0.35,\,\ell=2,\,m=-2,2$. It is interesting to notice that, for the real part of the prograde coefficients, some frequency domain estimates lie slightly outside the error bars. This effect can be ascribed to mode mixing, which becomes more prominent for large angular momentum. In fact, for the case $a=0.25$ shown in Fig.~\ref{fig:d_l2m2_a05} the time- and frequency-domain estimates are fully compatible. Moreover, the agreement for retrograde modes is typically better than the prograde case.

\begin{figure*}[ht]
\includegraphics[width=\linewidth]{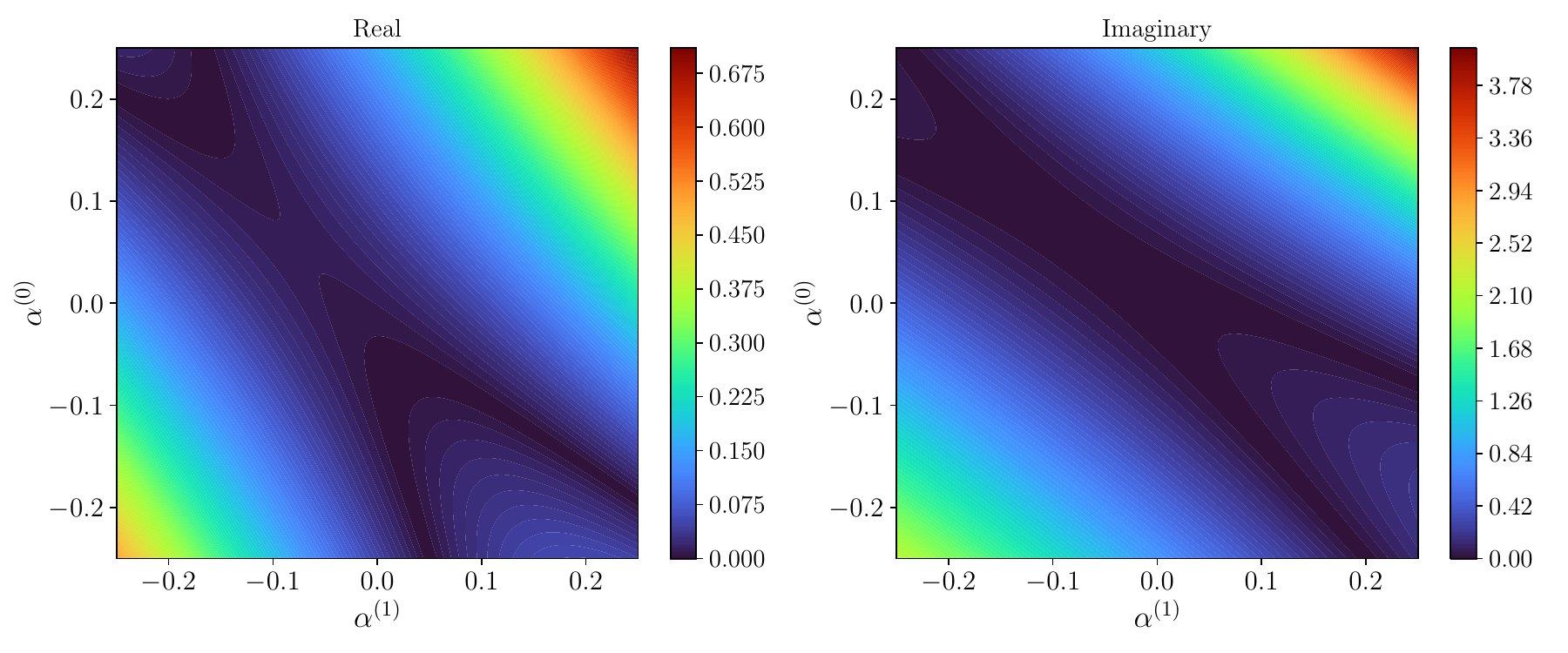}
\centering
\caption{Color map of the relative error ($\%$) between the linear and quadratic fit of the real and imaginary parts of the $a=0.25\,,\ell=2,\,m=0$ QNMs for the mixed coefficients $k=0$ and $k=1$ as a function of $\alpha^{(0)}$ and $\alpha^{(1)}$.}
\label{fig:color_map_mixed}
\end{figure*}

In Fig.~\ref{fig:color_map_mixed}, we report the relative error between the linear and quadratic fit obtained from Eq.~\eqref{higher_order_ParTeu}. The deviation from the linear approximation has the maximum value along the main diagonal of the  $(\alpha^{(0)},\alpha^{(1)})$ plane $\alpha^{(0)}=\alpha^{(1)}$; while it is minimal along the secondary diagonal of the plane $\alpha^{(0)}=-\alpha^{(1)}$. This type of behavior is observed for all pairs of indices $(k_1,k_2)$ in the interval $[-2,2]$.

\end{document}